# Deep Indentation of Soft Solids: A Limit State

Mohammad Shojaeifard[1], Zhenwei Ma[2], Jimmy Hsia[3], Norman Fleck[4], Mattia Bacca[1*]

[1]Mechanical Engineering, University of British Columbia, Vancouver, BC V6T 1Z4, Canada
[2]Surgery, University of British Columbia, Vancouver, V5Z 1M9, Canada
[3]Mechanical and Aerospace Engineering, Nanyang Technological University, 639798, Singapore
[4]Engineering Department, University of Cambridge, Trumpington St., Cambridge, CB2 1PZ, UK

[*]Corresponding author: mbacca@mech.ubc.ca

**Abstract**

Indentation is a practical technique for probing the mechanical properties of soft materials such as elastomers and gels when standard tests are difficult, destructive, or simply cannot be performed in-situ. Here, deep indentation of a half-space made from an isotropic, incompressible one-term Ogden hyperelastic solid is analysed for an indenter in the form of a rigid circular rod of radius $R$ and that is either flat-bottomed or has a hemispherical tip. Numerical simulations reveal that a limit state is achieved when the indent depth $D$ much exceeds the radius $R$ of the rod and the indentation force $F$ scales as $F = \beta E D^2$, where $\beta$ is a dimensionless coefficient and $E$ is the Young's modulus of the solid. The magnitude of $\beta$ increases with the degree of strain stiffening of the hyperelastic solid but is almost insensitive to the headshape of the rod and to the level of Coulomb friction at the interface with the solid. A high level of strain stiffening of the Ogden solid at finite strain implies that the level of strain at the indenter tip is small up to a large value of $D/R$, thereby protecting the solid from failure and penetration. Directly ahead of the indenter tip the contours of strain energy density are approximately spheroidal in shape and far ahead the Boussinesq solution is recovered. The limit state of deep indentation is compared to that of cavity expansion; whilst several qualitative features are shared, the two limit states differ in their kinematic fields and in their sensitivity to the degree of strain stiffening in the Ogden solid.

## 1. Introduction

Accurate mechanical characterization of soft materials exhibiting nonlinear hyperelastic behavior is essential for applications in biomedical engineering, soft matter physics, soft robotics, and materials design [Budday2020, Joodaki2018, Marechal2021]. Standard mechanical tests, most notably uniaxial tension, remain widely used for constitutive characterization, but they are often ill-suited for biological tissues and delicate soft materials because they require destructive specimen extraction, cannot be performed in situ or in vivo, and are prone to inaccuracies due to grip slippage [Jiang2020, Scholze2020]. Bulk compression tests, such as parallel-plate or unconfined compression, mitigate the problem of tensile gripping but still require sample extraction and preparation. In addition, specimen–platen friction alters the measured force–deformation response that must be characterized [Wu2004, Miller2005]. These limitations motivate the development of mechanical characterization methods that can be performed in situ or in vivo, do not require sample extraction, and are less sensitive to inaccuracies associated with contact friction.

Indentation-based methods provide a practical alternative because they are minimally invasive, require little or no specimen preparation, and can be applied directly to intact soft materials and biological tissues [Navindaran2023, He2024]. However, conventional indentation is usually in the shallow-depth regime, where the indentation depth $D$ is small compared with the contact radius. In this regime, the response is controlled by local contact geometry and is used to determine the small-strain elastic (Young's) modulus $E$ and the closely related shear modulus $\mu$, where $\mu = E/3$ for an incompressible, isotropic solid. For example, flat punch indentation of an incompressible elastomer by a flat-bottomed circular cylinder of radius $R$ generates an indentation force $F = (8/3)ER\,D$ according to Sneddon theory [Sneddon1965], while a spherically tipped indenter of tip radius $R$ generates a force $F = (16/9)ER^{0.5}D^{1.5}$ according to Hertz theory [Hertz1882]. These shallow-depth solutions allow for measurement of the small-strain elastic modulus $E$, but they do not identify the large-deformation hyperelastic response. Shallow indentation methods may also require knowledge of sample thickness and degree of heterogeneity, geometry and boundary conditions [Finan2014, Delaine-Smith2016].

Several studies have extended indentation beyond small-strain elastic characterization such that hyperelastic constitutive models are deduced from measured nonlinear force–depth responses. For example, spherical indentation has been combined with finite-element simulations and inverse analysis to estimate material parameters in neo-Hookean, Mooney–Rivlin, Fung, and Arruda–Boyce models [Zhang2014]. More recently, geometric-mapping approaches have proposed scaling laws for deep spherical indentation beyond the Hertzian regime, showing that geometric nonlinearity can dominate over material nonlinearity for common elastomers [Mu2026]. However, these approaches are tied to spherical indenter geometry and specific contact configurations, and do not establish whether a common deep-indentation response emerges from different indenter tip shapes.

In the present study, the incompressible one-term Ogden [Ogden1972] model is chosen to give a compact description of nonlinear hyperelasticity through a small-strain elastic (Young's) modulus $E$ and a strain-stiffening parameter $\alpha$, defined in the usual manner as follows. Write the strain energy density per unit volume $\Psi$ of an isotropic, incompressible solid as a function of the 3 principal stretches $(\lambda_1, \lambda_2, \lambda_3)$ such that

$$\Psi = \frac{2\mu}{\alpha^2}(\lambda_1^\alpha + \lambda_2^\alpha + \lambda_3^\alpha - 3) \tag{1}$$

in terms of a shear modulus $\mu = E/3$, and a parameter $\alpha$. Incompressibility implies immediately that $\lambda_1\lambda_2\lambda_3 = 1$. Multiple loading modes are often used for reliable calibration of $\alpha$ [Lohr2022]. Deep indentation may offer a practical route for identifying the nonlinear strain-stiffening response from a single indentation test.

Soft solids of high toughness have the feature that they can undergo large reversible indentations by a rod-like indenter prior to failure and subsequent penetration. This phenomenon has been investigated by Fakhouri et al (2015) [Fakhouri2015] and by Shojaeifard and Bacca (2025) [Shojaeifard2025]. Fakhouri et al (2015) [Fakhouri2015] performed deep indentation tests on acrylic triblock copolymer gel by means of a circular rod with either a flat bottom or a hemispherical tip. They observed that, prior to the onset of penetration by material tearing, the indentation force $F$ scales with the square of indentation depth $D$ such that $F/ED^2$ equals a constant, regardless of choice of head-shape. This force versus depth scaling law was attributed to the feature that their gel has a neo-Hookean uniaxial response; an assumed value of representative stretch at the indenter tip was used to justify this argument. However, no detailed finite element simulations were performed to determine the local stretch near the indenter tip, and to explore its relevance given the fact that the solid is almost incompressible and so hydrostatic stress must also be considered. Recently, Shojaeifard and Bacca (2025) [Shojaeifard2025] performed deep indentation tests on 3 silicone rubbers and on pig skin using a hemispherically tipped circular rod. The uniaxial responses of these 4 solids are far from neo-Hookean and are more accurately described by a one-term Ogden solid with large values of strain stiffening exponent, $\alpha$. Consistent with the observations of Fakhouri et al (2015) [Fakhouri2015], Shojaeifard and Bacca (2025) also observed a deep indentation regime such that the indentation force $F$ scales as $F/ED^2 = \beta$. They additionally found that the value of $\beta$ increases slightly with increasing $\alpha$. Thus, they obtained the remarkable result that $F$ scales with $D^2$ and is independent of the radius $R$ of the rod, regardless of choice of soft solid. Shojaeifard and Bacca (2025) [Shojaeifard2025] did not explore the effect of headshape on $F$ but did find that $F$ increases slightly with the level of interfacial friction between indenter and solid.

A major aim of the present study is to explore the physical nature of the limit state in deep indentation associated with the characteristic response $F/ED^2 = \beta$ . To do so, deep indentation tests are performed on the three silicone rubbers and pig skin of Shojaeifard and Bacca (2025) [Shojaeifard2025] using a flat-bottomed indenter; these experiments augment the previous indentation tests by Shojaeifard and Bacca (2025) [Shojaeifard2025] using a spherically-tipped indenter. Detailed finite element simulations of indentation are performed for both flat-bottomed and spherically tipped indenters by making use of the incompressible one-term Ogden hyperelastic model [Ogden1972] as stated in (1). Finite element simulations are used to quantify the expansion with indentation depth of an activated region, defined as the region in which the strain energy density exceeds 1% of the shear modulus, while photoelasticity experiments provide an independent visualization of the associated expansion of the deformed region.

Further, the limit state of deep indentation is also compared qualitatively and quantitatively with the limit state of spherical cavity expansion in an infinite solid under an internal pressure. Recall that the cavity expansion solution for internal pressurisation of a spherical cavity in an infinite solid has been used as an approximate solution of the indentation pressure, see for example Johnson (1987) [Johnson1987]. However, the relevance of cavitation to indentation is questioned as the far-field solution for spherical cavity expansion differs from the Boussinesq field for a point load on an elastic half-space. Also, the near field solutions for cavity expansion and for indentation differ. Thus, there is no a priori reason that cavity expansion should be relevant to indentation. Rather, cavitation of a spherical cavity and deep indentation are two examples of a limit state associated with the finite deformation of a solid under different loadings.

## 2. Materials characterisation and deep indentation tests

### 2.1 Uniaxial response

The solids considered in this study are the same as those used in a previous study by Shojaeifard and Bacca (2025) [Shojaeifard2025]: three commercially available silicone-based elastomers, Ecoflex 00-10, Ecoflex 00-30, and Mold Star 16 Fast (all supplied by Smooth-On Inc); and porcine skin. The uniaxial tension responses for the 4 solids were also obtained by [Shojaeifard2025] and full experimental details are reported there. Here we repeat only the results in Figure 1, along with the curve fits using the Ogden solid (1). The faint grey lines show the large-stretch asymptotic scaling on log–log axes, while the solid curves show the one-term Ogden uniaxial response for the uniaxial nominal stress $S = 2E\left(\lambda^{\alpha-1} - \lambda^{-\alpha/2-1}\right)/(3\alpha)$ in terms of the uniaxial stretch $\lambda$. Tabulated values for $E$ and $\alpha$ are given in Table 1.

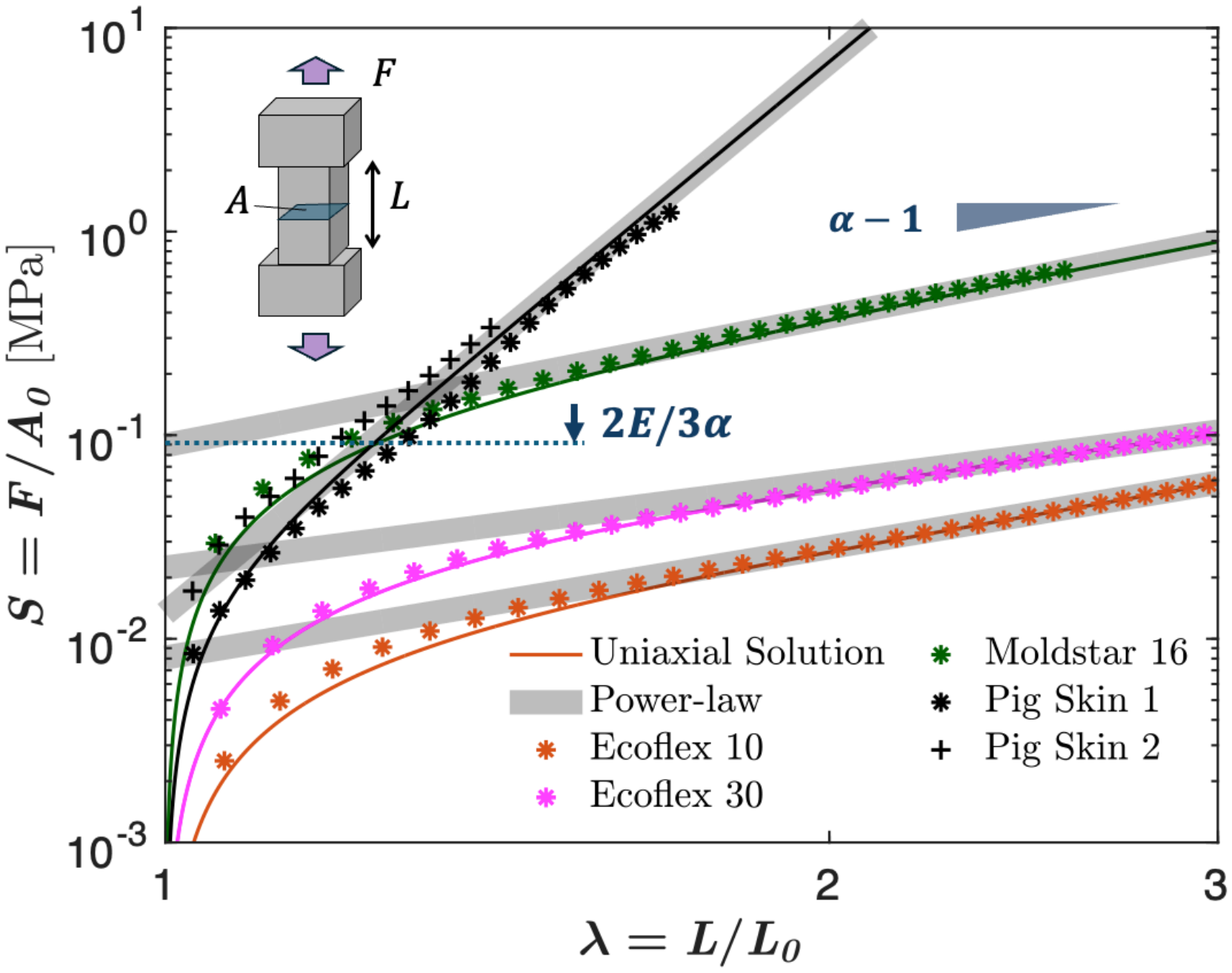

**Figure 1**: Uniaxial tensile tests reported in [Shojaeifard2025] for a gauge section of initial length $L_0$ and initial cross-sectional area $A_0$, highlighting the strain-stiffening coefficient $\alpha$. At large stretch, the nominal stress S follows the asymptotic relation $S \approx \lambda^{\alpha-1} 2E/(3\alpha)$; thus, on log–log axes, the slope approaches $\alpha - 1$, and extrapolation to $\lambda = 1$ gives $2E/(3\alpha)$, from which the modulus $E$ can be extracted.

**Table 1**: Comparison between material parameters identified from indentation experiments and reference values obtained from uniaxial tension tests reported in [Shojaeifard2025]. Errors are reported separately for flat-ended and spherical-tipped indentation relative to the uniaxial reference values.

| | ***Uniaxial tension*** [Shojaeifard2025] | ***Error flat indenter vs uniaxial*** | ***Error spherical indenter vs uniaxial*** |
|---|---|---|---|

| | $E_U$ $[kPa]$ | $\alpha_U$ | $e_{E_{f-u}}$ | $e_{\alpha_{f-u}}$ | $e_{E_{s-u}}$ | $e_{\alpha_{s-u}}$ |
|---|---|---|---|---|---|---|
| **Ecoflex 10** | 34.0 | 2.8 | 11.7% | 7.2% | 10.9% | 8.3% |
| **Ecoflex 30** | 81.0 | 2.4 | -1.4% | 11.1% | -0.74% | 10.1% |
| **Mold Star 16** | 420 | 3.1 | -1.07% | 6.7% | 1.79% | 2.8% |
| **Pig Skin** | 280 | 10 | -14.2% | -8.8% | -22.7% | -1.9% |

## 2.2 Deep Indentation Tests

The effect of headshape on the deep indentation response of the 3 silicone rubbers and pig skin was determined by performing indentation tests using flat-bottomed cylindrical stainless-steel rods of radius $R = 0.5$ mm. The results for indenters with a hemispherical tip of radius 0.5 mm are taken from [Shojaeifard2025], while those for flat tip are generated following the same sample preparation. Tests were performed using a screw-driven universal testing machine under displacement control at a constant cross-head speed of $0.1$ $\mathrm{mm/s}$, with the punch displacement given by the cross-head displacement and the axial load measured by the load cell of the test machine. The specimens comprised circular cylinders of radius $40$ mm and height $45$ mm, both much larger than the indenter radius. No lubricant was used during the tests.

The silicone elastomers were cast into 3D-printed molds. For porcine skin, we removed the subcutaneous fat to obtain a layer of uniform thickness of 3 mm; the skin was cut into circular disks of diameter 80 mm, and 14 layers were then stacked and bonded together with cyanoacrylate adhesive to attain the desired specimen height of 42 mm. The skin samples were kept hydrated in phosphate-buffered saline throughout preparation and testing. These materials span a broad range of stiffness and strain-stiffening behaviors.

The indentation depths were up to $D$ = 8 mm, corresponding to normalized indentation depths up to $D/R = 16$, implying that all tests extended well beyond the shallow linear-elastic regime. Force–displacement data were recorded at 5 $Hz$ during loading and unloading. For each elastomer and indenter geometry, we performed three repeat tests; deviations between repeated measurements remained below 4%, and the resulting curves were averaged. As shown previously in [Shojaeifard2025], interfacial friction has only a mild influence on the deep-indentation response, particularly for strongly strain-stiffening materials.

Figure 2 contains a plot of $F/D^2$ versus $D/R$ on log-log axes for each of the 4 soft solids. For each choice of solid, as the deep-indentation regime is approached, $D \gg R$, $F/D^2$ converges to a constant value that scales with the Young's modulus,

$$\frac{F}{D^2} = \beta E \tag{2}$$

Here, $\beta$ is a dimensionless coefficient that depends upon the hyperelastic response of the solid. It is clear from Fig. 2 that the value of $\beta$ is independent of headshape of indenter. But, in order to extract the value of $\beta$ from deep indentation tests, it is necessary to first make use of the shallow indentation response in order to deduce a value for the Young's modulus $E$.

For a spherical indenter and an incompressible sample, the classical Hertz solution [Hertz1882] for shallow indentation gives

$$\frac{F}{D^2} = \frac{16}{9} E \left(\frac{D}{R}\right)^{-0.5} \quad (3)$$

The data in Fig. 2 for shallow indentation by the spherically tipped indenter support this: the slope equals $-0.5$. Alternatively, for a flat circular punch indenting an incompressible solid, the shallow indentation elastic response is given by the Sneddon solution [Sneddon1965] as $F = (8/3)ER\,D$, which is rephrased to

$$\frac{F}{D^2} = \frac{8}{3} E \left(\frac{D}{R}\right)^{-1} \quad (4)$$

The slope of $-1$ in Figure 2a supports this. Faint reference lines indicate the shallow-indentation linear-elastic limits: spherical indentation follows the Hertzian scaling $F/D^2 \sim D^{-0.5}$, with intercept $16E/9$ at $D = R$, while flat-ended indentation follows $F/D^2 \sim D^{-1}$, with intercept $8E/3$ at $D = R$.

Extrapolation of the shallow linear-elastic fits to the point $D = R$ in Fig. 2 gives $F/D^2 = 16E/9$ for spherical indentation from (3) and $F/D^2 = 8E/3$ for flat-punch indentation from (4). These intercepts provide the small-strain elastic modulus $E$. After values of $E$ have been determined for each solid, the horizontal asymptote reached at large depth, $D \gg R$, is used with (2) to extract the coefficient $\beta$. The parameters extracted from this two-regime asymptotic fitting procedure are reported in Table 2 for Ecoflex 00-10, Ecoflex 00-30, Mold Star 16 Fast, and porcine skin. Table 1 compares these values with parameters obtained from uniaxial tension tests reported in [Shojaeifard2025].

To describe the indentation response across regimes, we adopt

$$\frac{F}{D^2} = E\left[\beta + \frac{16}{9}\left(\frac{D}{R}\right)^{-0.5} \Theta\left(\frac{D}{R}\right)\right] \quad (5)$$

for spherical indenters, and

$$\frac{F}{D^2} = E\left[\beta + \frac{8}{3}\left(\frac{D}{R}\right)^{-1} \Theta\left(\frac{D}{R}\right)\right] \quad (6)$$

for flat indenters, where the transition function is taken to be

$$\Theta\left(\frac{D}{R}\right) = \exp\left(-\gamma \frac{D}{R}\right)$$

and controls the attenuation of the shallow linear-elastic contribution with increasing indentation depth. Here, $\gamma$ is an attenuation coefficient. We take $\gamma = \beta$ for spherical tips and $\gamma = 0.75\,\beta$ for flat tips, which provides good agreement with the experimental shallow-to-deep transition shown by the model curves in Fig. 2.

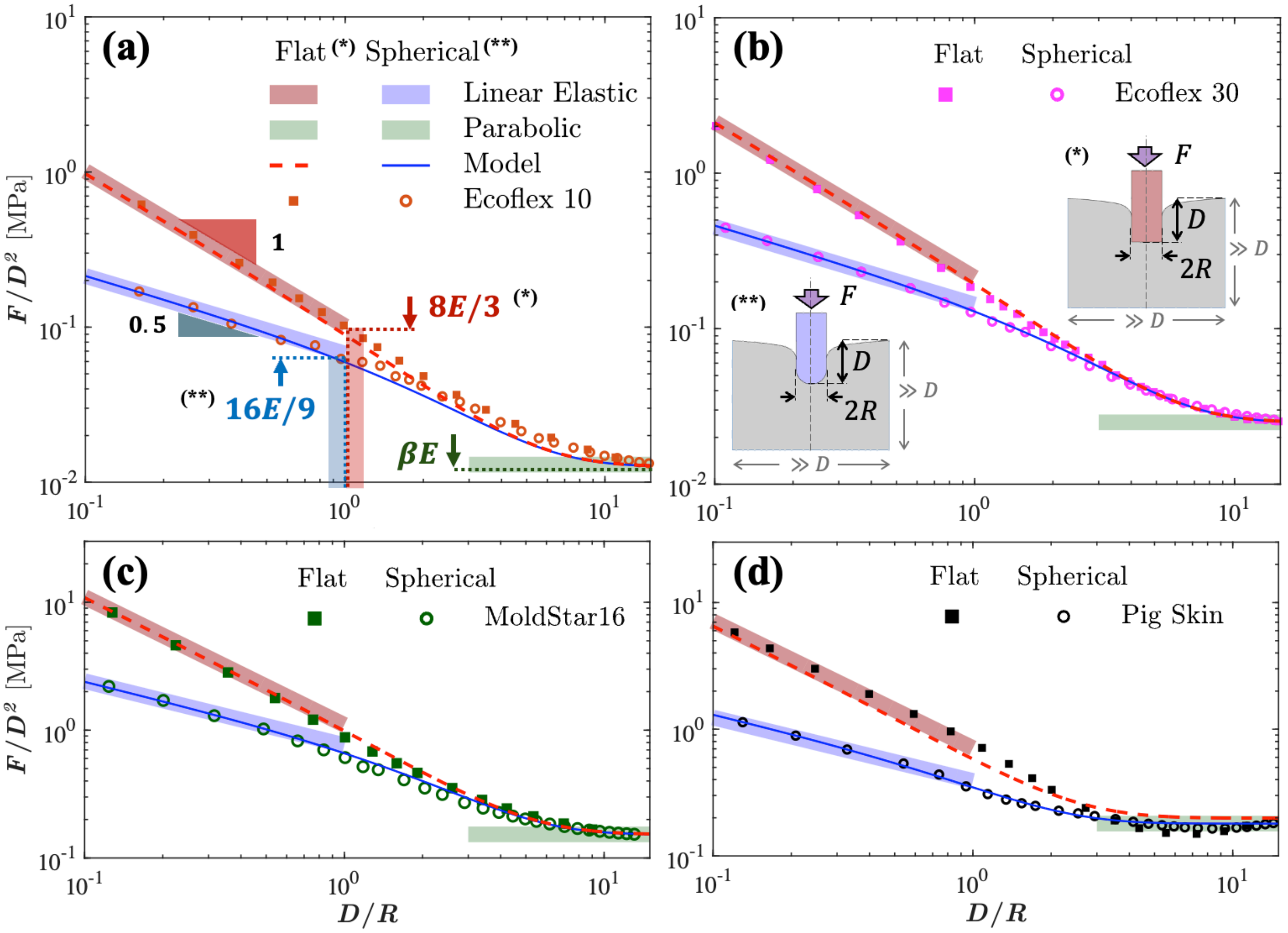

**Figure 2**: Experimental force–depth response for (a) Ecoflex® 00-10, (b) Ecoflex® 00-30, (c) Mold Star™ 16, and (d) porcine skin. Normalized force $F/D^2$ is plotted against normalized indentation depth $D/R$ for flat-ended and spherical-tipped cylindrical indenters. Symbols denote experimental data, while dashed and solid curves denote the corresponding theoretical predictions for flat-ended and spherical indenters, respectively, Eqs. (5)-(6). Insets in (b) illustrate the two indenter geometries and the shallow- and deep-indentation regimes, $D \ll R$ and $D \gg R$.

Tables 1 and 2 also report discrepancies in estimates of the parameter $E$ (modulus) and $\alpha$ (strain stiffening) between indenter shapes and between indentation and uniaxial testing. We use subscripts $u$, $f$, and $s$ to denote parameters obtained from uniaxial tension, flat-ended indentation, and spherical-tipped indentation, respectively. In Table 1, errors are computed using uniaxial tension as the reference, with $e_{E(fu)} = \left(E_f - E_u\right)/E_u$, $e_{\alpha(fu)} = \left(\alpha_f - \alpha_u\right)/\alpha_u$, $e_{E(su)} = (E_s - E_u)/E_u$, and $e_{\alpha(su)} = (\alpha_s - \alpha_u)/\alpha_u$. In Table 2, errors are computed using spherical indentation as the reference, with $e_E = \left(E_f - E_s\right)/E_s$ and $e_\alpha = (\alpha_f - \alpha_s)/\alpha_s$. For the silicone elastomers, the discrepancies between flat and spherical indentation, and between indentation and uniaxial tension, are about 10% or less. Larger deviations are observed for porcine skin, with errors on the order of 20%, consistent with the greater heterogeneity, anisotropy, and layered structure of biological tissue. In addition, the porcine skin data show a mild deviation from the ideal deep-indentation plateau predicted by Eq. (2), suggesting that intrinsic material or structural length scales may affect the response at large indentation depths. This point is revisited in Section 6.

**Table 2**: Material parameters extracted from indentation experiments using flat-ended and spherical-tipped indenters. The small-strain elastic modulus $E$ was obtained from the shallow linear-elastic regime, the parabolic coefficient $\beta$ from the deep-indentation plateau, and the Ogden strain-stiffening parameter $\alpha$ by inverting the $\beta(\alpha)$ relation (Fig. 3). The relative differences $e_E$ and $e_\alpha$ quantify the discrepancy between parameters obtained from flat and spherical indentation.

| | ***Flat tip Indentation*** | | | ***Spherical tip Indentation*** | | | ***Error*** | |
|---|---|---|---|---|---|---|---|---|
| | $E_f$ $[kPa]$ | $\beta_f$ | $\alpha_f$ | $E_s$ $[kPa]$ | $\beta_s$ | $\alpha_s$ | $e_{E_{f-s}}$ | $e_{\alpha_{f-s}}$ |
| **Ecoflex 10** | 38.0 | 0.342 | 3.001 | 37.7 | 0.345 | 3.035 | 0.79% | -1.13% |
| **Ecoflex 30** | 79.9 | 0.313 | 2.666 | 80.4 | 0.311 | 2.642 | -0.62% | 0.92% |
| **Mold Star 16** | 415.5 | 0.368 | 3.309 | 427.5 | 0.358 | 3.188 | -2.8% | 3.8% |
| **Pig Skin** | 240.2 | 0.79 | 9.120 | 216.5 | 0.831 | 9.81 | 10.9% | -7% |

## 3. Finite Element Analysis

The commercial finite element code Abaqus/Explicit 2024 was used to analyze the transition from shallow linear-elastic indentation to the deep indentation regime and to determine the dependence of the coefficient $\beta$ upon the degree of hyperelastic strain stiffening. A two-dimensional axisymmetric model was used of a rigid indenter and a circular cylindrical soft solid. We considered both flat-ended and spherical-tipped cylindrical indenters of unit radius $R$. To approximate a half-space response, a circular cylindrical sample was considered with radius and height equal to $100R$. The domain was discretized into approximately 100,000 hybrid axisymmetric elements with reduced integration and hourglass control, and the rigid indenter was defined by axisymmetric rigid elements. The indentation displacement was prescribed and the indentation velocity was sufficiently small to maintain quasi-static conditions, with the kinetic-to-internal energy ratio kept below 2%. Contact between the indenter and soft solid was idealised by Coulomb friction with friction coefficient $f = \tau_f/p$, where $\tau_f$ is the frictional shear strength, and $p$ is the contact pressure. To quantify the sensitivity of the deep-indentation coefficient $\beta$ to interfacial friction, simulations were performed for $f = 0$, 0.1, and 1, as reported in the right panel of Fig. 3. Additional simulations with different friction coefficients produced similar trends, especially for strain-stiffening materials, confirming the findings of [Shojaeifard2025]. Unless otherwise stated, the force–depth simulations reported here use $f = 0.3$.

Predictions of non-dimensional force $F/(ED^2)$ versus non-dimensional indentation depth $D/R$ are shown in Fig. 3 for flat-ended and spherical-tipped indenters, using $\alpha = 2,3,5,9$. At shallow depths, the simulations recover the corresponding linear-elastic limits: the spherical indenter follows Hertzian scaling, while the flat-ended indenter follows the Sneddon flat-punch scaling. At larger depths, both responses transition toward the plateau value $F/(ED^2) = \beta$, highlighting the parabolic force–depth law. For $\alpha = 3{,}5{,}9$, the flat and spherical indentation curves converge to the same deep-indentation plateau, confirming that $\beta$ is independent of local tip shape in this regime, and showing that $\beta$ increases with $\alpha$.

We extracted $\beta$ from the plateau value of $F/(ED^2)$ at large $D/R$ for several values of $\alpha$. The relation between these two parameters is shown in the right panel of Fig. 3 and is described by the empirical fit

$$\beta \approx 0.15\, \alpha^{0.75} \tag{7}$$

provided by [Shojaeifard2025].

This correlation is used to infer a value of α from the measured value of $\beta$ for each of the 4 soft solids, recall Fig. 2. This protocol was used to obtain the values of α reported in Table 2.

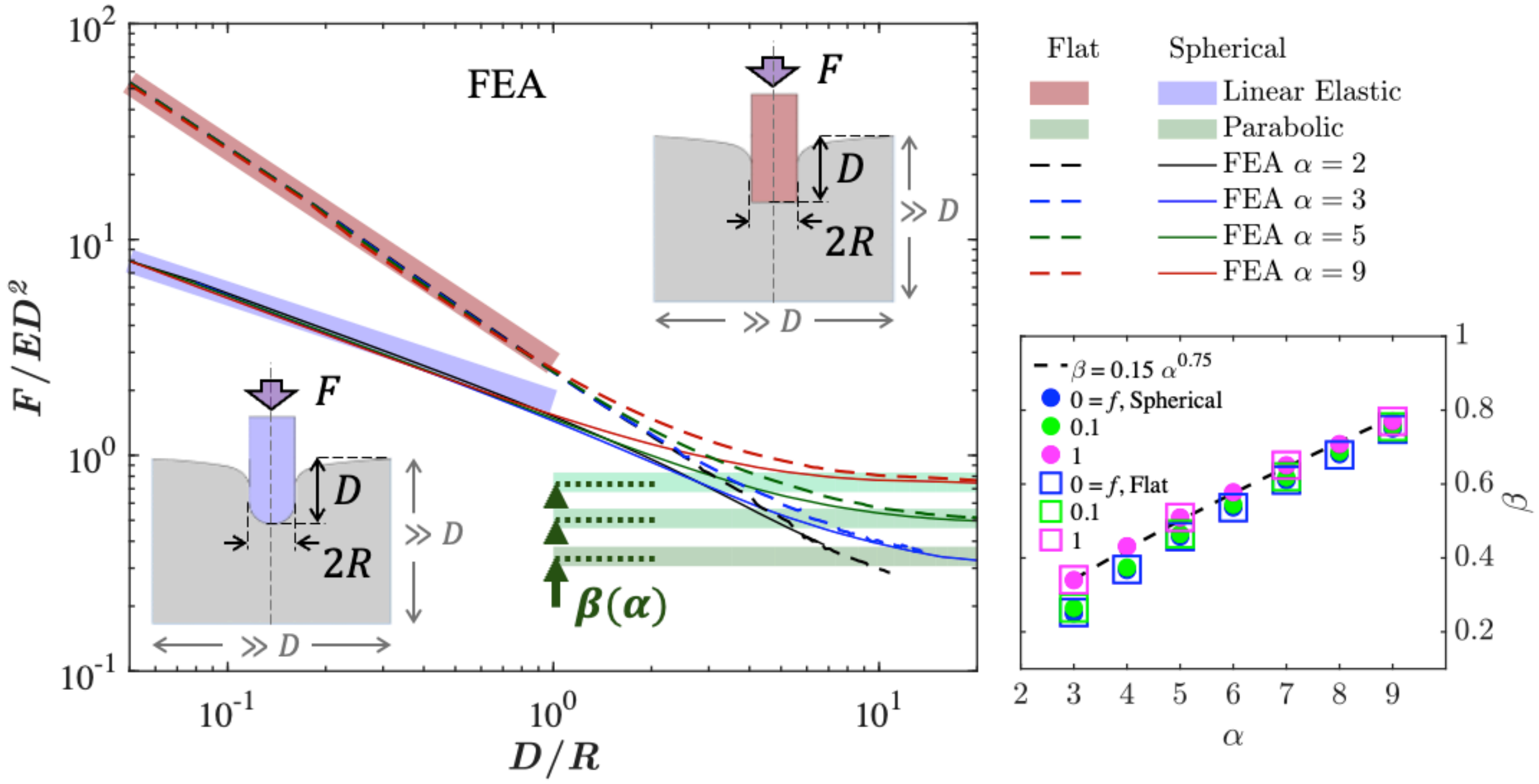

**Figure 3**: Finite-element results for flat-ended and spherical-tipped cylindrical indentation of incompressible one-term Ogden solids with strain-stiffening parameter $\alpha$. Left: normalized force $F/(ED^2)$ versus normalized indentation depth $D/R$. Shaded bands indicate the linear-elastic and deep-indentation parabolic regimes; dashed and solid curves denote flat and spherical indentation, respectively. Right: deep-indentation coefficient $\beta$ as a function of $\alpha$ for both indenter geometries, together with the empirical fit $\beta \approx 0.15\, \alpha^{0.75}$ provided by [Shojaeifard2025].

## 4. The Limit State of Deep Indentation

It is instructive to use the finite element simulations to interrogate the nature of the deep indentation limit state such that $F/ED^2$ is a constant. First, consider the distribution of strain energy density (SED) in the vicinity of the spherically tipped indenter. A representative solution is shown in Fig.

4(b) for the choice $\alpha = 9$ and $D/R = 30$. We identify a so-called 'mechanically activated zone' as the portion of the sample where the strain energy density (SED) exceeds 1% of the shear modulus $\mu$. As shown in Fig. 4, this region forms a spheroidal domain around the indenter. We calculated its width $W$ and height $H$ in the deformed configuration, and the corresponding dimensions $W_0$ and $H_0$ in the undeformed configuration. The undeformed dimensions $W_0$ and $H_0$ increase approximately linearly with indentation depth at large $D/R$, for both flat-ended and spherical-tipped indenters.

The rate of this growth also increases with the strain-stiffening parameter $\alpha$. Since the characteristic dimensions of the mechanically active region scale with $D$, its volume scales as $D^3$. Further, the average strain energy density in this region is constant since $D \gg R$; consequently, the total stored elastic energy scales as $U \sim ED^3$. Upon differentiating $U$ with respect to indentation depth the indentation force follows as $F = dU/dD \sim ED^2$, which explains the parabolic force–depth response.

The spheroidal activated region growth is also shown in the preliminary photoelastic experiments in Fig. 5, generated by indenting a PDMS sample prepared with a 5:1 base-to-curing-agent ratio, having Young's modulus $E = 3.1\ MPa$ and strain-stiffening parameter $\alpha = 4$. The isochromatic fringes give contours of principal stress difference due to suitable use of cross-polaroids. The contours expand with increasing indentation depth, reaching a characteristic diameter of about $2D$, consistent with the FEA results.

We further characterize the spatial extent of this deformation field by noting that the only relevant dimensional parameters in the problem are the deep indentation distance $D$ (when much larger than $R$), the force $F$ and the Young's modulus $E$. Consequently, dimensional analysis demands that $F/ED^2$ is a constant. It is clear from Fig. 3 that $F/ED^2 = \beta$ is on the order of unity. Hence the characteristic length scale $a = \sqrt{F/E}$ is on the order of $D$.

Figure 6 shows the normalized strain-energy-density profiles along the radial direction $r$ at $z = 0$ and along the axial direction $z$ at $r = 0$ at 3 selected values of $D/R$ = 1, 5 and 10. Distances are normalized by the characteristic length $a = \sqrt{F/E}$ for the three values of indentation force $F$ associated with the abovementioned values of $D/R$. Far from the contact, the finite-element profiles approach the Boussinesq far-field solution for a normal point force acting on an elastic half-space [Boussinesq1885, Johnson1987]. The characteristic length $a = \sqrt{F/E}$ is chosen for this comparison because the Boussinesq solution contains no geometric length scale associated with the indenter; the only quantities defining a length scale are the applied point force $F$ and elastic modulus $E$ of the indented material.

Along the free surface, $z = 0$, the Boussinesq solution gives

$$\frac{SED}{\mu} = \frac{(1+\nu)^2(1-2\nu)^2}{2\pi^2}\left(\frac{a}{r}\right)^4 \tag{8a}$$

while along the loading axis, at $r = 0$, the Boussinesq solution gives

$$\frac{SED}{\mu} = \frac{(1+\nu)^2(19-12\nu-4\nu^2)}{8\pi^2}\left(\frac{a}{z}\right)^4 \tag{8b}$$

In the case of an incompressible material like the elastomers analyzed here, the energy density in Eq. (8a) would vanish, due to $\nu = 0.5$. Thus, for the Boussinesq comparison, we evaluate Eq. (8) at $\nu = 0.47$, giving

$$\frac{SED}{\mu} = 3.94 \cdot 10^{-4} \left(\frac{a}{r}\right)^4 \tag{9a}$$

$$\frac{SED}{\mu} = 0.34 \left(\frac{a}{z}\right)^4 \tag{9b}$$

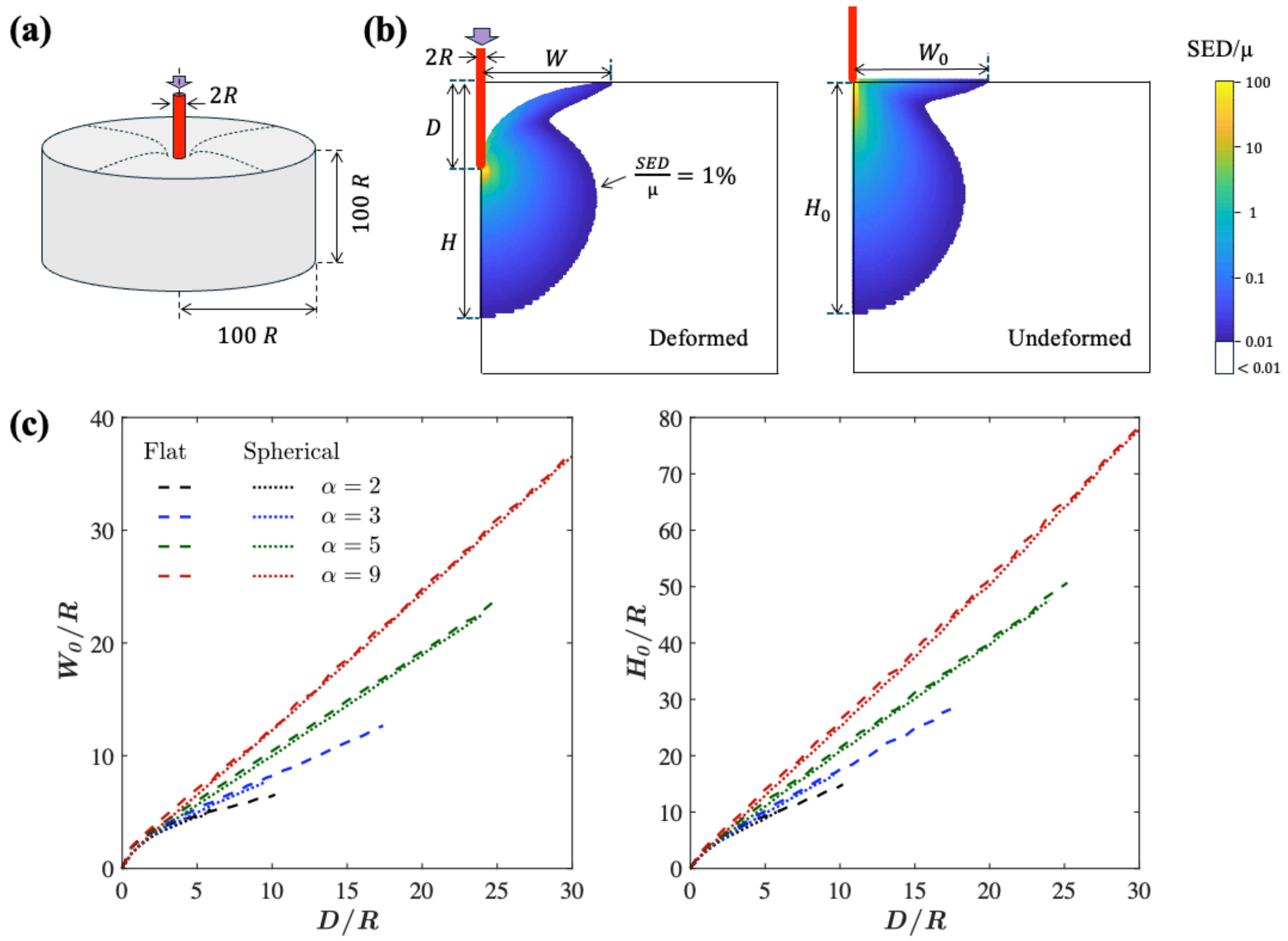

**Figure 4**: Growth of the mechanically active region during deep indentation. (a) Finite-element geometry for a cylindrical sample of radius and height $100R$ indented by a rigid cylinder of radius $R$. (b) Mechanically active region, defined by $SED/\mu > 0.01$, shown in the deformed and undeformed configurations for the choice $\alpha = 9$; $W$, $H$ and $W_0$, $H_0$ denote its corresponding dimensions. (c) $W_0/R$ and $H_0/R$ versus $D/R$ for flat-ended and spherical-tipped indenters and different Ogden parameters $\alpha$.

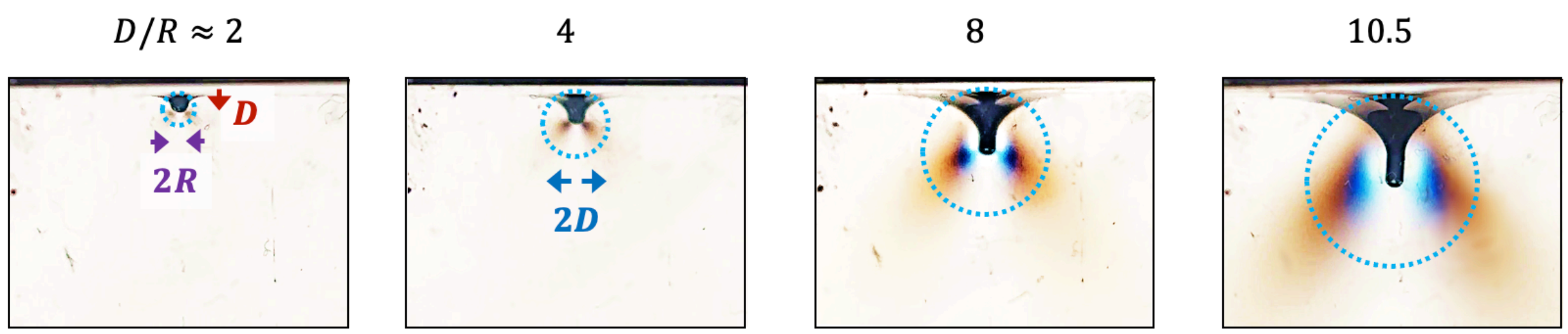

**Figure 5**: Photoelastic images during deep indentation at increasing $D/R$. The birefringent region expands with indentation depth, with a characteristic diameter of about $2D$.

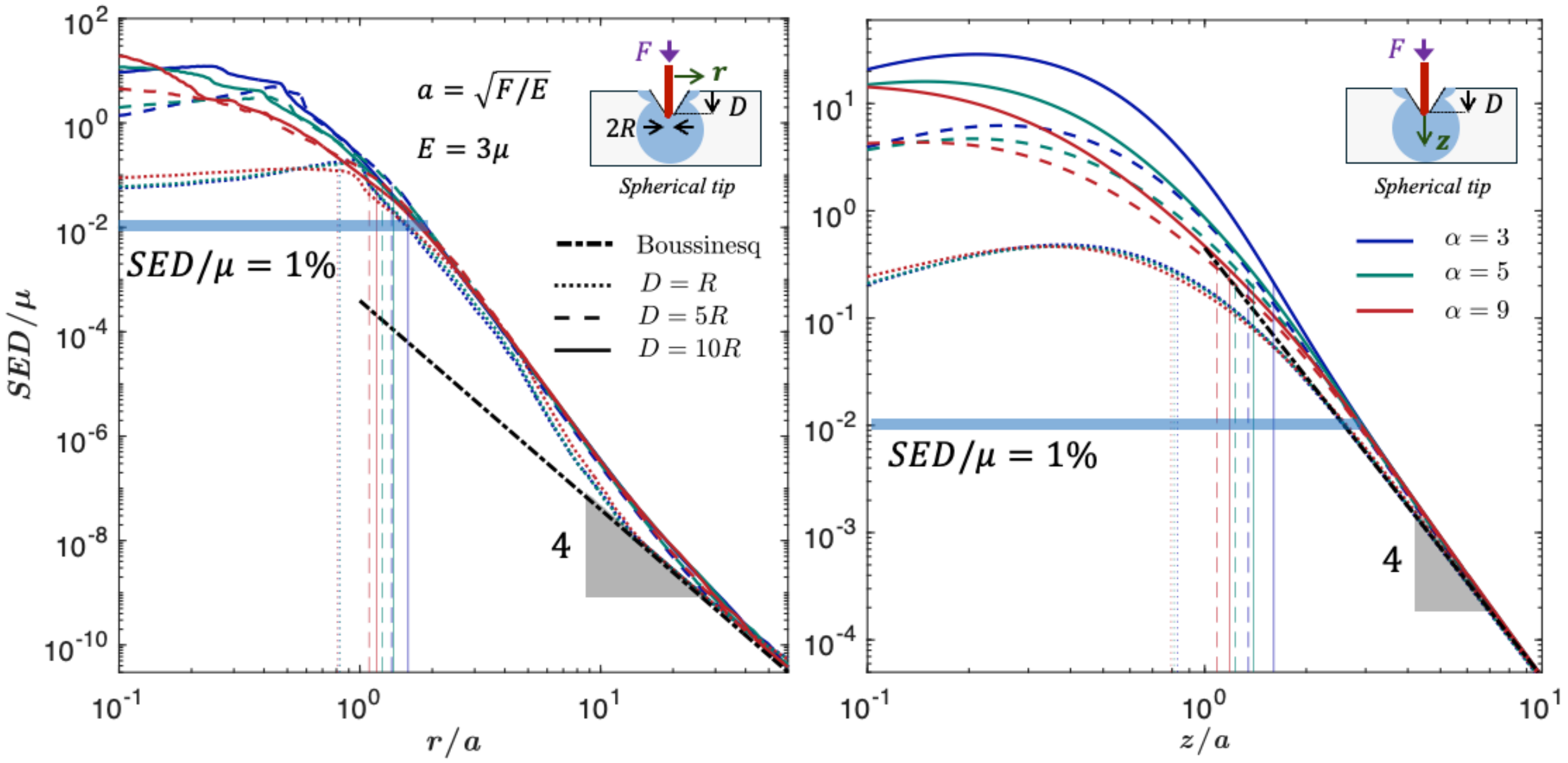

**Figure 6**: Strain-energy-density (SED) profiles from deep-indentation FEA. Normalized $SED/\mu$ is plotted against normalized radial distance $r/a$ at $z = 0$ (left) and axial distance $z/a$ at $r = 0$ (right), with $a = \sqrt{F/E}$. Results are shown for different indentation depths $D/R$ and Ogden parameters $\alpha$. The horizontal line denotes the threshold $SED/\mu = 1\%$, and the black dash-dotted curve denotes the Boussinesq far-field solution. The colored vertical lines indicate the normalized position corresponding to one indentation depth, $r/a = D/a$ on the free surface (left) and $z/a = D/a$ along the loading axis (right), and are included only as a reference for comparing the spatial extent of the activated region with the indentation depth.

## 5. Comparison of limit state of deep indentation with that for cavitation

Deep indentation has previously been studied in the context of elastoplasticity, and in early models the indentation pressure has been idealised by the cavitation pressure for cavity expansion [Bishop1945, Johnson1970]. In the deep penetration of an elastoplastic solid, such as a metal, the deformation fields around the indenter resemble those associated with cylindrical and spherical cavity expansion. This analogy motivates a comparison between deep indentation and cavity expansion in hyperelastic solids.

When a cavity in a hyperelastic solid is internally pressurized, it may attain a limit state of continued cavity growth at constant internal pressure. For some choices of solid this limiting cavitation pressure is finite, whereas for others it is infinite. As shown in *Appendix A* for an incompressible one-term Ogden solid, a finite limiting pressure exists for $\alpha < 2$ in cylindrical cavity expansion and for $\alpha < 3$ in spherical cavity expansion, whereas the pressure is unbounded for $\alpha$ at or above these respective thresholds. When a finite cavitation-pressure limit is attained, the pressure remains constant while the stretch at the cavity wall continues to increase as the cavity enlarges.

Deep indentation also attains a limit state with continued indentation such that $F/D^2$ approaches a finite value, while the local stretch at the indenter tip continues to increase with indentation depth. Figure 7 compares the evolution of the maximum local stretch $\lambda_{max}$ in the two problems. In the spherical cavity expansion problem, $\lambda_{max} = a/a_0$ is the hoop stretch at the inner wall of the cavity after the cavity has expanded from an initial radius $a_0$ to a current radius $a$ (*Appendix A*). For spherical-tip indentation, $\lambda_{max}$ corresponds to the equal radial and circumferential stretches at the axis of symmetry immediately ahead of the indenter tip. In both indentation and cavity expansion, $\lambda_{max}$ continues to increase with the corresponding deformation measures $D/R$ and $a/a_0 - 1$. However, the rate of increase differs between the two problems: for cavity expansion, $\lambda_{max}$ increases linearly with $a/a_0 - 1$, and is independent of the material parameter $\alpha$; by contrast, during deep indentation, the evolution of $\lambda_{max}$ with $D/R$ depends strongly on $\alpha$ as this parameter controls the distribution of stretch in the indented solid. This demonstrates that the cavity-expansion analogy established for deep indentation in elastoplastic materials does not extend to hyperelastic solids that undergo strain stiffening.

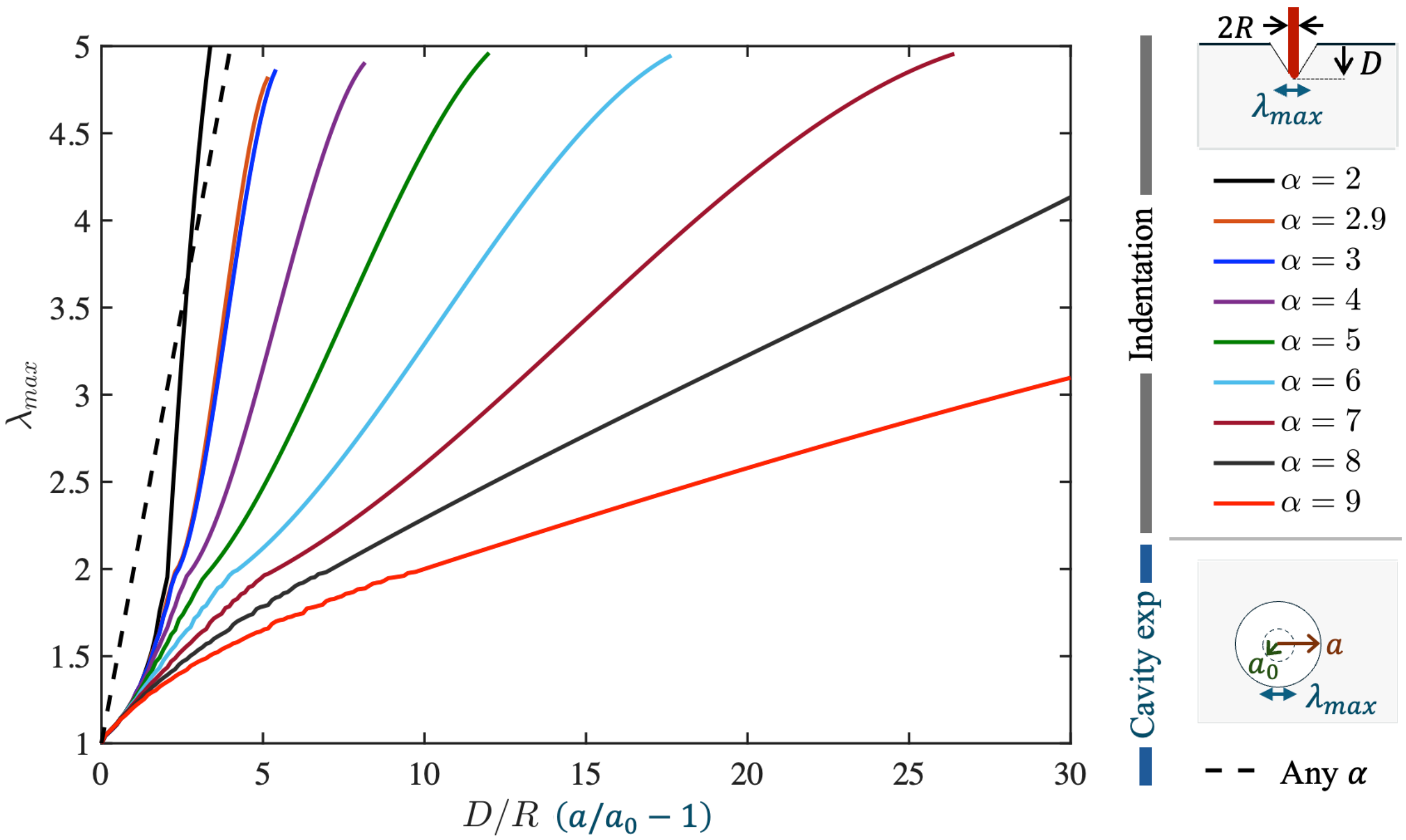

**Figure 7**: Comparison of the maximum stretch during deep indentation and cavity expansion. For spherical-tip indentation, the maximum stretch $\lambda_{max}$ immediately ahead of the indenter is plotted against normalized indentation depth $D/R$ for different Ogden strain-stiffening parameters $\alpha$. For cavity expansion, $\lambda_{max} = a/a_0$ is plotted against $a/a_0 - 1$, giving the universal dashed line independent of $\alpha$. In both cases, the maximum local stretch continues to increase with deformation, but at different rates.

## 6. Limitations of deep indentation theory

Mechanisms of puncture or cutting [Fakhouri2015, Fregonese2021, Fregonese2022, Fregonese2023, Goda2024a, Goda2024b, Wu2026] terminate or can precede the parabolic deep-indentation force-depth regime. Thus, hyperelastic characterization by deep indentation requires an experimental window in which the indentation depth is sufficiently large to satisfy $D \gg R$, but smaller than the critical puncture depth. For sharp or cylindrical indenters, previous studies have shown that this critical depth scales as $D_c \sim \sqrt{\ell_e R}$, where $\ell_e = \Gamma/\mu$ is the elasto-fracture length (critical crack opening radius upon crack propagation), $\Gamma$ is the fracture toughness, and $\mu$ is the shear modulus [Fakhouri2015, Wu2026]. This scaling implies that a nondestructive deep-indentation window is most readily obtained when the indenter radius is smaller than, or at most comparable to, the elasto-fracture length $\ell_e$. Consequently, soft and tough materials, which have large $\ell_e$, can be probed using larger indenters, whereas stiff or brittle materials may require much smaller indenters. Such miniaturization introduces practical constraints, including reduced force resolution and possible indenter buckling; a full design criterion coupling puncture avoidance, force sensitivity, and indenter stability is an important step for future work.

A further assumption of the present theory is that the specimen behaves as a half-space, such that its lateral dimensions and thickness are large compared with the extent of the deformation field.

Finite specimen dimensions and proximity to external boundaries can modify the indentation and puncture response [Barney2024]; consequently, the present scaling is expected to apply when the activated region remains sufficiently far from the specimen boundaries.

Another assumption underlying the proposed theory is that indentation occurs under quasi-static conditions, so that inertial and rate-dependent effects are negligible. This imposes a constraint on the characteristic loading time of the experiment, $t_c = D/V$, where $V$ is the indentation velocity. Such time must remain much longer than the material relaxation time $t_r$, so that the material response can be considered the relaxed hyperelastic one. The quasi-static assumption therefore requires $t_c \gg t_r$, or equivalently $V \ll D/t_r$. Many biological tissues, however, exhibit pronounced rate-dependent behavior arising from intrinsic viscoelasticity due to interchain friction and/or poroelastic fluid transport. Extending the framework to deep viscoelastic indentation is an important direction for future work, where the apparent modulus $E$ and deep-indentation coefficient $\beta$ are expected to become rate dependent.

Further, the present theory assumes that the material is homogeneous. While this approximation is appropriate for many synthetic soft materials, such as the silicone elastomers we have tested, biological tissues, including the pig skin analyzed here, frequently exhibit layered architectures and spatially varying constitutive properties. The region of large straining expands progressively during deep indentation and consequently the measured response may represent an average of the constitutive behavior over an increasingly large volume, leading to deviations from the universal homogeneous scaling law presented here. This may partly explain the larger parameter-estimation errors observed for pig skin compared with the elastomers. In particular, Fig. 2d shows that the normalized force $F/D^2$ does not converge to a plateau but instead continues to increase with indentation depth. The non-parabolic scaling at deep indentation introduces an additional length scale associated with the multilayer structure of skin, for example. Extending the present framework to heterogeneous and layered materials, particularly for the case of a stiffer skin layer on the surface of the sample, will broaden the applicability of our method to a wider range of biological tissues, but will also introduce new variables, thus increasing the complexity of the characterization protocol.

## 7. Concluding remarks

Despite the limitations discussed above, this work establishes a universal deep-indentation regime for hyperelastic half-spaces in which the classical dependence on indenter tip size disappears. Unlike shallow contact mechanics, where the indenter radius and tip geometry strongly influence the response, sufficiently deep indentation follows a tip-independent parabolic force–depth relationship governed by the elastic modulus and the nonlinear constitutive response of the material. As indentation proceeds, a spheroidal region of activated material, characterized by non-negligible strain energy density, expands into the bulk and progressively dominates the mechanical response. Consistently, far from the contact region, the stress field resembles that of an elastic half-space loaded by a concentrated force, as described by the classical Boussinesq solution [Boussinesq1885, Johnson1987].

The present study suggests a systematic means for constitutive characterization. Lohr et al. [Lohr2022] emphasized that the one-term Ogden model is widely used for isotropic soft tissues because it describes nonlinear hyperelasticity using only two parameters: the small-strain shear modulus and the strain-stiffening parameter $\alpha$. But they also noted that the reliable identification of $\alpha$ is more challenging than measurement of the small-strain modulus, often requiring multiple loading modes or inverse finite-element analyses when a tensile stress versus strain curve is not available due to specimen size limitations. The present framework provides a complementary route. The shallow-indentation regime identifies the small-strain elastic modulus $E$ from the dependence of the normalized force $F/D^2$ on the dimensionless depth $D/R$, while the tip-independent deep-indentation plateau $\beta E$ determines $\beta$, which is uniquely related to the Ogden strain-stiffening parameter through (7). Consequently, both the small-strain modulus and the Ogden strain-stiffening parameter can be extracted from a single indentation experiment. This provides a simple, mechanically interpretable, and potentially minimally invasive strategy for characterizing soft materials and biological tissues.

In the deep-indentation regime, $D \gg R$, the deformation is no longer controlled by the headshape of the indenter. Instead, the applied force engages a spheroidal nonlinear-elastic (NLE) region, shown in blue in Fig. 8, where the strain energy density (SED) exceeds 1% of the shear modulus $\mu$. The size of this region of strain scales with the characteristic length $a = \sqrt{F/E}$. In the deep-indentation regime, $a$ is comparable in magnitude to the indent depth $D$; consequently, the volume of material that is subjected to finite strain, also known as the 'activated volume' scales as $V \sim D^3$. Suppose the average SED in this region is on the order of the Young's modulus $E$; then, the stored elastic energy scales as $U \sim ED^3$, and the parabolic force–depth response scales as $F \sim ED^2$. This energetic interpretation explains why, in the deep-indentation regime, the response becomes controlled by the growth of the activated bulk deformation field rather than by local tip geometry. Deep indentation emerges as a means of probing nonlinear hyperelasticity in a wide range of soft solids.

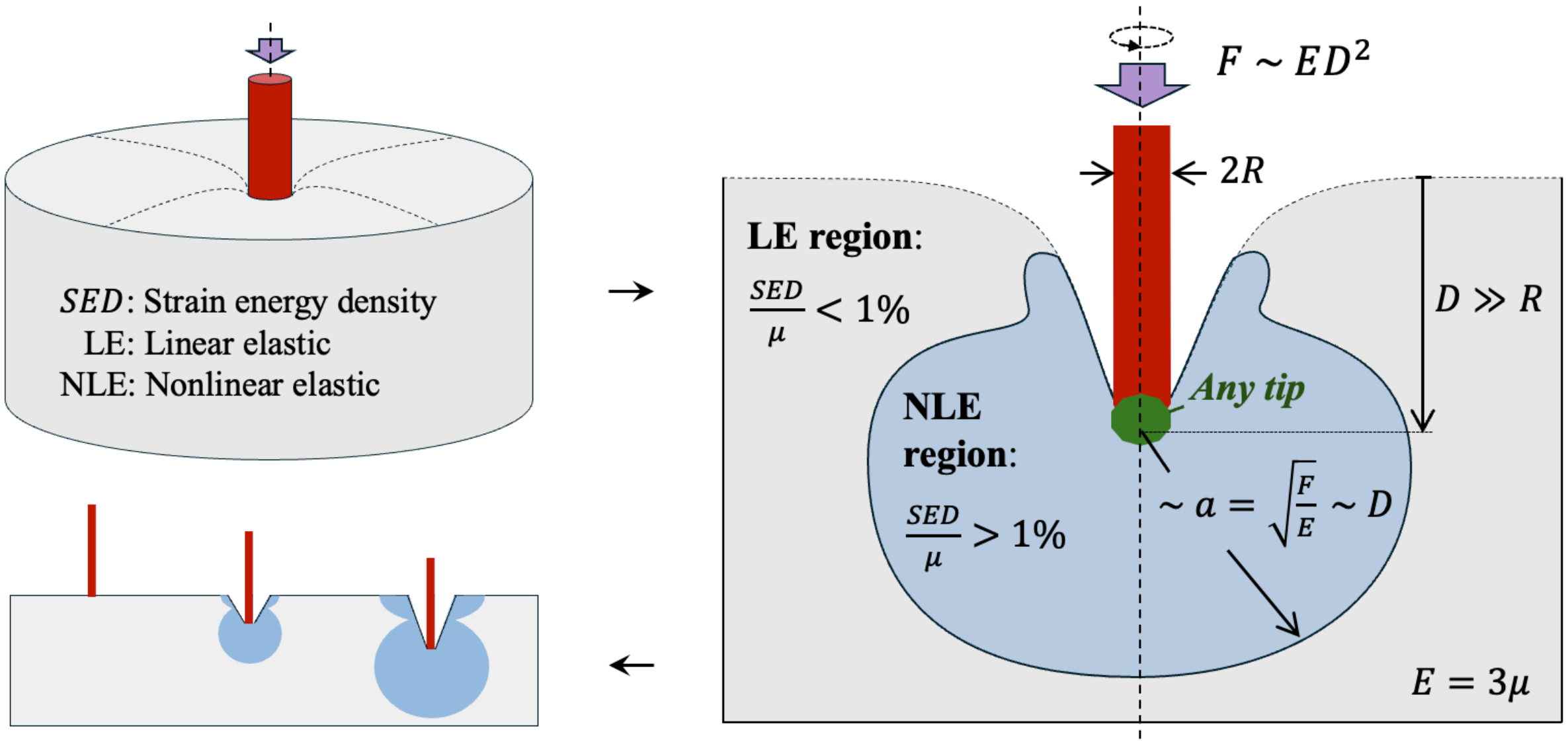

**Figure 8**: Schematic of the deep-indentation process. A soft hyperelastic sample is indented by a rigid cylindrical indenter of radius $R$, with a generic tip shape here represented schematically as either flat or spherical.

## Appendix A – Hyperelastic cavity expansion

The following derivation of the cavitation limit for internal pressurisation of an incompressible solid follows that of Huang et al (1991) [Huang1991] for a spherical cavity and Wright et al (1991) [Wright1991] for a circular cylindrical cavity. To describe the displacement field during cavity expansion in an incompressible solid, we use the general formulation

$$r^{m-1}\, dr = s^{m-1}\, ds \qquad \text{(A1)}$$

where $r$ and $s$ are the radial coordinates in the current and reference configuration, respectively; $m = 2$ for a cylindrical cavity under plane strain, and $m = 3$ for a spherical cavity. The inner wall of the cavity is identified by $r = a$ and $s = a_0$, so that $a$ and $a_0$ are the cavity radii in the current and reference configurations, respectively. The solution of Eq. (A1) is,

$$r^m = s^m + a^m - a_0^m \qquad \text{(A2)}$$

We then define the hoop stretch as

$$\lambda = \frac{r}{s} \qquad \text{(A3)}$$

So that at the cavity wall we have the maximum (hoop) stretch

$$\lambda_{max} = \frac{a}{a_0} \qquad \text{(A4)}$$

The radial stretch then follows from incompressibility as

$$\lambda_r = \lambda^{-(m-1)} \qquad \text{(A5)}$$

For the cylindrical problem the third principal stretch is unity, whereas for spherical expansion the two hoop stretches are equal to $\lambda$. The Ogden strain-energy density in (1) can therefore be written in the form

$$\frac{SED}{\mu} = \frac{2}{\alpha^2}\left[(m-1)\lambda^{\alpha} + \lambda^{-(m-1)\alpha} - m\right] \qquad \text{(A6)}$$

The principal Cauchy stresses are

$$\sigma_i = -p + \lambda_i \frac{\partial SED}{\partial \lambda_i} \qquad \text{(A7)}$$

where $p$ is the hydrostatic pressure associated with incompressibility. Consequently,

$$\sigma_\theta - \sigma_r = \frac{2\mu}{\alpha}\left(\lambda^{\alpha} - \lambda^{-(m-1)\alpha}\right) \qquad \text{(A8)}$$

Radial equilibrium requires

$$\frac{d\sigma_r}{dr} = \frac{m-1}{r}(\sigma_\theta - \sigma_r) \qquad \text{(A9)}$$

For a cavity subjected to an internal pressure $P$, we have

$$\sigma_r(a) = -P,\ \ \sigma_r(\infty) = 0 \qquad \text{(A10)}$$

and integration of the radial equilibrium in Eq. (A9) gives

$$P = \int_a^\infty \frac{m-1}{r}(\sigma_\theta - \sigma_r)\, dr \quad \text{(A11)}$$

Differentiation of $\lambda = r/s$, together with incompressibility, gives

$$\frac{ds}{s} = \frac{\lambda^{m-1}}{1-\lambda^m}\, d\lambda \quad \text{(A12a)}$$

or, equivalently,

$$\frac{dr}{r} = -\frac{d\lambda}{\lambda(\lambda^m - 1)} \quad \text{(A12b)}$$

Now substitute into (A11) to obtain the cavity pressure in terms of the cavity-wall stretch,

$$\frac{P}{\mu} = \frac{2(m-1)}{\alpha}\int_1^{\lambda_{max}} \frac{\lambda^\alpha - \lambda^{-(m-1)\alpha}}{\lambda(\lambda^m - 1)}\, d\lambda \quad \text{(A13)}$$

for either cylindrical or spherical expansion.

The existence of a limiting cavitation pressure follows directly from the large-$\lambda$ behavior of this expression. For $\lambda \gg 1$,

$$\frac{\lambda^\alpha - \lambda^{-(m-1)\alpha}}{\lambda(\lambda^m - 1)} \sim \lambda^{\alpha-m-1} \quad \text{(A14)}$$

and hence

$$\frac{P}{\mu} \sim \frac{2(m-1)}{\alpha}\int_1^{\lambda_{max}} \lambda^{\alpha-m-1} d\lambda \quad \text{(A15)}$$

It follows that

$$\begin{aligned} &\alpha < m: \quad P \to P_c < \infty \\ &\alpha = m: \quad \frac{P}{\mu} \sim \frac{2(m-1)}{m}\ln(\lambda_{max}) \to \infty \\ &\alpha > m: \quad \frac{P}{\mu} \sim \frac{2(m-1)}{\alpha(\alpha-m)}\lambda_{max}^{\alpha-m} \to \infty \end{aligned} \quad \text{(A16)}$$

Thus, the existence of a finite limiting pressure depends both on the material strain-stiffening parameter and on the cavity geometry. Specifically, a finite limiting pressure requires $\alpha < 2$ for cylindrical cavity expansion ($m = 2$) and $\alpha < 3$ for spherical cavity expansion ($m = 3$).

**Acknowledgments**

This work was supported by the Natural Sciences and Engineering Research Council of Canada (NSERC) (RGPIN-2025-07085).